\documentstyle[aps,graphicx,multicol]{revtex}

\draft

\begin{document}

\title{Free energy and criticality in the nucleon pair breaking process}
\author{M.~Guttormsen\footnote{Electronic address: magne.guttormsen@fys.uio.no}, R.~Chankova, M.~Hjorth-Jensen, J.~Rekstad, and S.~Siem}
\address{Department of Physics, University of Oslo, N-0316 Oslo, Norway}
\author{A.~Schiller}
\address{Lawrence Livermore National Laboratory, L-414, 7000 East Avenue,
Livermore, California 94551, U.S.A.}
\author{D.J.~Dean}
\address{Physics Division, Oak Ridge National Laboratory, P.O. Box 
2008, Oak Ridge, Tennessee 37831, U.S.A.}
\maketitle

\begin{abstract}
Experimental level densities for $^{171,172}$Yb, $^{166,167}$Er, 
$^{161,162}$Dy, and $^{148,149}$Sm are analyzed within the microcanonical 
ensemble. In the even isotopes at excitation energies $E<2$~MeV, the Helmholtz 
free energy $F$ signals the transition from zero to two quasiparticles. For 
$E>2$~MeV, the odd and even isotopes reveal a surprisingly constant $F$ at a 
critical temperature $T_c\sim 0.5$~MeV, indicating the continuous melting of 
nucleon Cooper pairs as function of excitation energy.
\end{abstract}


\begin{multicols}{2}

\section{Introduction}

One of the most spectacular pairing phase transitions in nature is the 
transition from a normal to a superconducting phase in large electron systems. 
The transition is triggered at low temperature by massive pairing of two and 
two electrons into spin $J=0$ pairs, so called Cooper pairs\cite{BC57}.

For atomic nuclei, the pairing phase transition is expected to behave 
differently. First of all, the nucleus is an isolated, few body system with two
species of fermions. Surface effects are prominent, and the coherence length of
nucleons coupled in Cooper pairs is larger than the nuclear diameter. 
Furthermore, there are non-negligible energy spacings between the single 
particle orbitals. All these facts make the nucleus an inherently small system.
Also, other types of residual interactions than pairing are of importance. The 
influence of these peculiar constraints on the nucleus has been investigated 
theoretically for a long time \cite{SY63,Mo72,TS80,Go81}, however, only limited
experimental information is available to describe the nature of pairing within 
the nucleus.

The Oslo group has developed a method to derive simultaneously the level 
density and $\gamma$-ray strength function from a set of primary $\gamma$-ray 
spectra \cite{SB00}. The method has been well tested and today a consistent 
data set for eight rare earth nuclei is available. In the present work we 
report for the first time on a comprehensive analysis of the evolution of the 
pairing phase transition as a function of the nuclear excitation energy.

\section{Experimental level densities}

Level densities for $^{171,172}$Yb, $^{166,167}$Er, $^{161,162}$Dy, and 
$^{148,149}$Sm have been extracted from particle-$\gamma$ coincidences. The 
experiments were carried out with 45~MeV $^3$He projectiles accelerated by the
MC-35 cyclotron at the University of Oslo. The data were recorded with the 
CACTUS multidetector array using the pick up ($^3$He,$\alpha\gamma$) reaction 
on $^{172,173}$Yb, $^{167}$Er, $^{162,163}$Dy, and $^{149}$Sm targets and the 
inelastic ($^3$He,$^3$He'$\gamma$) reaction on $^{167}$Er and $^{149}$Sm 
targets. The charged ejectiles were detected with eight $\Delta E$--$E$ 
particle telescopes placed at an angle of 45$^{\circ}$ relative to the beam 
direction. Each telescope comprises one Si front and one Si(Li) back detector 
with thicknesses of 140 and 3000~$\mu$m, respectively. An array of 28 NaI 
$\gamma$-ray detectors with a total efficiency of $\sim$15\% surrounds the 
target and particle detectors. From the reaction kinematics, the measured 
ejectile energy can be transformed into excitation energy $E$. Thus, each 
coincident $\gamma$ ray can be assigned to a $\gamma$ cascade originating from 
a specific energy $E$. These spectra are the basis for the extraction of level 
density and $\gamma$-strength function as described in Ref.\ \cite{SB00}. 
Several interesting applications of the method have been demonstrated, see, 
e.g., Refs.\ \cite{SB01,MG01,VG01,SG02}.

\begin{figure}
\includegraphics[totalheight=20cm,angle=0,bb=0 20 350 730]{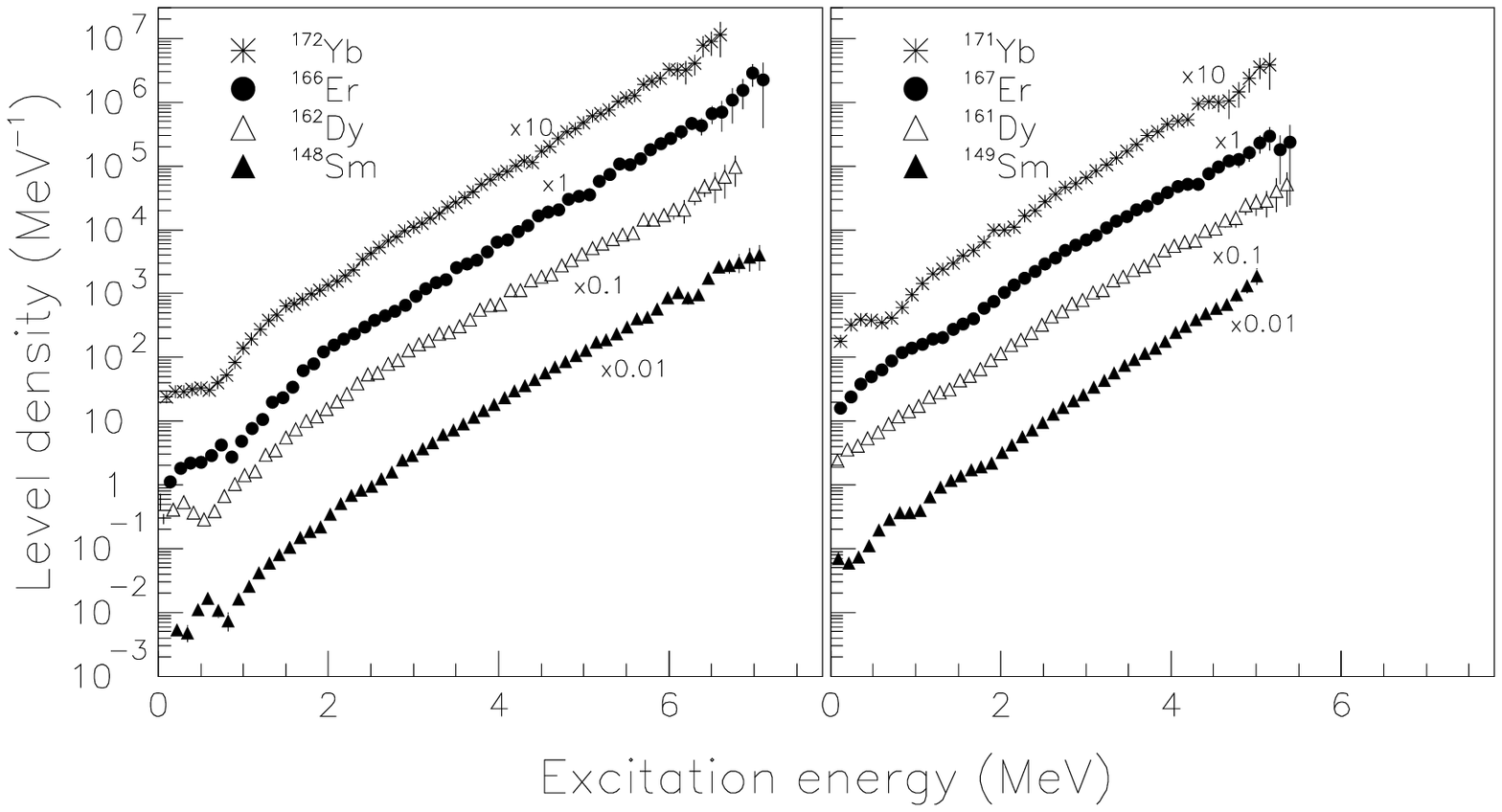}
\caption{Experimental level densities for the nuclei $^{171,172}$Yb, 
$^{166,167}$Er, $^{161,162}$Dy, and $^{148,149}$Sm. The data are taken from 
Refs.\ \protect\cite{SB01,MG01,SG02}.}
\label{fig:fig1}
\end{figure}

The level densities for $^{171,172}$Yb, $^{166,167}$Er, $^{161,162}$Dy, and 
$^{148,149}$Sm are shown in Fig.\ \ref{fig:fig1}. The level densities are 
normalized at low excitation energies where (almost) all levels are known, and 
at the neutron binding energy $B_n$ where the level density can be estimated 
from neutron-resonance spacings. The spin window populated in the reactions is 
typically $I\sim 2$--$6\hbar$. Already, three general comments can be made to 
these data: (\em i\rm) above 2~MeV excitation energy, all level densities are 
very linear in a log plot, suggesting a so-called constant temperature level 
density, (\em ii\rm) the level densities of the odd-even isotopes are larger 
than for their neighboring even-even isotopes, and (\em iii\rm) the even-even 
isotopes show a strong increase in level density between 1 and 2~MeV, 
indicating the breaking of Cooper pairs.

It should be noted that the transitions considered here are low temperature 
phenomena. The $^{171,172}$Yb, $^{166,167}$Er, and $^{161,162}$Dy nuclei have 
well deformed shapes, and various calculations in this mass region 
\cite{DK93,WK00,ER00,AS01} indicate that the transition from deformed to 
spherical shape occurs at much higher temperatures than the temperature at 
which the first pairs break. However, for nuclei closer to the $N=82$ shell 
gap, e.g., $^{148,149}$Sm, the coexistence between deformed and spherical 
shapes at low temperatures cannot be excluded, as discussed in Ref.\ 
\cite{ZC99}.

\section{Free energy and critical temperature}

The statistical microcanonical ensemble is an appropriate working frame for 
describing an isolated system like the nucleus. In this ensemble the excitation
energy $E$ is fixed, in accordance with the observables of our experiments. The
microcanonical entropy is given by the number of levels $\Omega$ at $E$
\begin{equation}
S(E)=k_B\ln\Omega(E),
\label{eq:S(E)}
\end{equation}
where the multiplicity $\Omega$ is directly proportional to the level density 
$\rho$ by $\Omega(E)=\rho(E)/\rho_0$. The normalization denominator $\rho_0$ is
adjusted to give $S\sim 0$ for $T \sim 0$, which fulfills the third law of 
thermodynamics. Here, we assume that the lowest levels of the ground state 
bands of the $^{172}$Yb, $^{166}$Er, $^{162}$Dy, and $^{148}$Sm nuclei have 
temperatures close to zero, giving on the average $\rho_0=2.2$~MeV$^{-1}$. In 
the following, this value is used for all eight nuclei and Boltzmann's constant
is set to unity ($k_B=1$).

In order to analyze the criticality of low temperature transitions, we 
investigate the probability $P$ of a system at the fixed temperature $T$ to 
have the excitation energy $E$, i.e.,
\begin{equation}
P(E,T,L)=\Omega(E)\exp\left(-E/T\right)/Z(T),
\end{equation}
where the canonical partition function is given by 
$Z(T)=\int_0^\infty\Omega(E')\exp\left(-E'/T\right)\,dE'$. Implicitly, the 
multiplicity of states $\Omega(E)$ depends on the size of the system, denoted 
by $L$. Often, it is more practical to use the negative logarithm of this 
probability $A(E,T)=-\ln P(E,T)$, where we in the following omit the $L$ 
parameter. Lee and Kosterlitz showed \cite{LK90,LK91} that the function 
$A(E,T)$, for a fixed temperature $T$ in the vicinity of a critical temperature
$T_c$ of a structural transition, will exhibit a characteristic double-minimum 
structure at energies $E_1$ and $E_2$. For the critical temperature $T_c$, one 
finds $A(E_1,T_c)=A(E_2,T_c)$. It can be easily shown that $A$ is closely 
connected to the Helmholtz free energy and the previous condition is equivalent
to
\begin{equation}
F_c(E_1)=F_c(E_2),
\end{equation}
a condition which can be evaluated directly from our experimental data. Here, 
it should be emphasized that $F_c$ is a linearized approximation to the 
Helmholtz free energy at the critical temperature $T_c$ according to 
$F_c(E)=E-T_cS(E)$, thereby avoiding the introduction of a caloric curve 
$T(E)$. The free-energy barrier at the intermediate energy $E_m$ between $E_1$ 
and $E_2$, is given by
\begin{equation}
\Delta F_c=F_c(E_m)-F_c(E_1).
\end{equation}
Now, the evolution of $\Delta F_c$ with increasing system size $L$ may 
determine the order of a possible phase transition \cite{LK90,LK91}. These 
ideas have, e.g., recently been applied to analyze phase transitions in a 
schematic pairing model \cite{BD01}.

Figure \ref{fig:fig2} displays a schematic description of the entropy for 
even-even, odd-mass and odd-odd nuclei as function of excitation energy. In the
lower excitation energy region of the even-even nucleus, only the ground state 
is present, and above $E\sim 2\Delta$ the level density is assumed to follow a 
constant temperature formula. It has been shown \cite{GB00,GH01} that the 
single particle entropy $s$ is an approximately extensive thermodynamical 
quantity in nuclei at these temperatures. The increase in entropy at the 
breaking of the first proton or neutron pair, i.e., at $E=2\Delta$, is roughly
$2s$ in total for the two newly created unpaired nucleons. The requirement 
$F_c(E_1)=F_c(E_2)$, where $T=T_c$, gives 
$E_2-E_1=T_c\left[S(E_2)-S(E_1)\right]$. Thus, with the assumed estimates 
above, we obtain the relation $\Delta=T_cs$, which may be used to extract the 
critical temperature for the pairing transition. Adopting typical values of 
$\Delta=1$~MeV and $s=2k_B$ \cite{GB00,GH01}, we obtain $T_c=0.5$~MeV\@. For 
the odd-mass case, one starts out with one quasiparticle which gives roughly 
one unit of single-particle entropy $s$ around the ground state. The three 
quasiparticle regime appears roughly at $E=2\Delta$ with a total entropy of 
$3s$. The region between $E=0$ and 2~MeV is modeled with a step at 
$\sim 2$~MeV, however, in real nuclei the level density is almost linear in a 
log-plot for the whole excitation energy region due to the smearing effects of 
the valence nucleon. In the case of an odd-odd nucleus, one starts out with two
units of single-particle entropy. The two valence nucleons represent a strong 
smearing effect on the level density and the modeled step structure in entropy 
at $E=2\Delta$ for the onset of the four quasiparticle regime is completely 
washed out.

In the higher excitation region, further steps for transitions to higher 
quasiparticle regimes are also washed out due to the strong smearing effects of
the already present, unpaired nucleons. The slope of the entropy with 
excitation energy is determined by two competing effects: the quenching of 
pairing correlations which drives the cost in energy lower for the breaking of 
additional pairs, and the Pauli blocking which reduces the entropy created per 
additional broken pair. The competing influence of both effects is modeled by a
constant-temperature level density with the same slope for all three nuclear 
systems and a slightly higher critical temperature. In this region of the 
model, there are infinitely many excitation energies where the relation 
$F_c(E_1)=F_c(E_2)$ is fulfilled.

The breaking of proton or neutron pairs are thought to take place at similar 
excitation energies due to the approximate isospin symmetry of the strong 
interaction. It is indeed commonly believed that the pairing gap parameter 
$\Delta$ and, thus, the critical temperature $T_c$ for the breaking of Cooper 
pairs, are approximately the same for protons and neutrons. Furthermore, 
interactions between protons and neutrons will certainly wash out any 
differences in behavior between the proton and neutron fluids. In Fig.\ 
\ref{fig:fig3}, the influence of differences in proton and neutron pair 
breaking is investigated within our schematic model. Here, we assume that 
neutrons break up at $2\Delta$ creating an entropy of $2s$. The protons are 
assumed to break up at 10\% higher excitation energy (since $Z<N$) creating 
10\% less entropy (due to the larger proton single particle level spacing). The
entropy of the total system of either proton or neutron pair breaking gives
\begin{equation}
S(E)=\ln\left[e^{S_p(E)}+e^{S_n(E)}\right]-\ln2,
\end{equation}
where the last term assures that $S=0$ in the ground state band. The 
requirement $F_c(E_1)=F_c(E_2)$ gives $T_c^{(n)}=\Delta/s$ for neutrons (as in 
Fig.\ \ref{fig:fig2}) and $T_c^{(p)}=1.1\Delta/0.9s=1.22\Delta/s$ for protons. 
In the combined system of both neutrons and protons, a value of 
$T_c=(T_c^{(n)}+T_c^{(p)})/2=1.11\Delta/s$ is deduced from Fig.\ 
\ref{fig:fig3}. Thus, typical fluctuations in the pairing gap parameter and the
single particle entropy for neutrons and protons give only small changes in the
extracted critical temperature.

\section{Experimental results}

\begin{figure}
\includegraphics[totalheight=20cm,angle=0,bb=0 20 350 730]{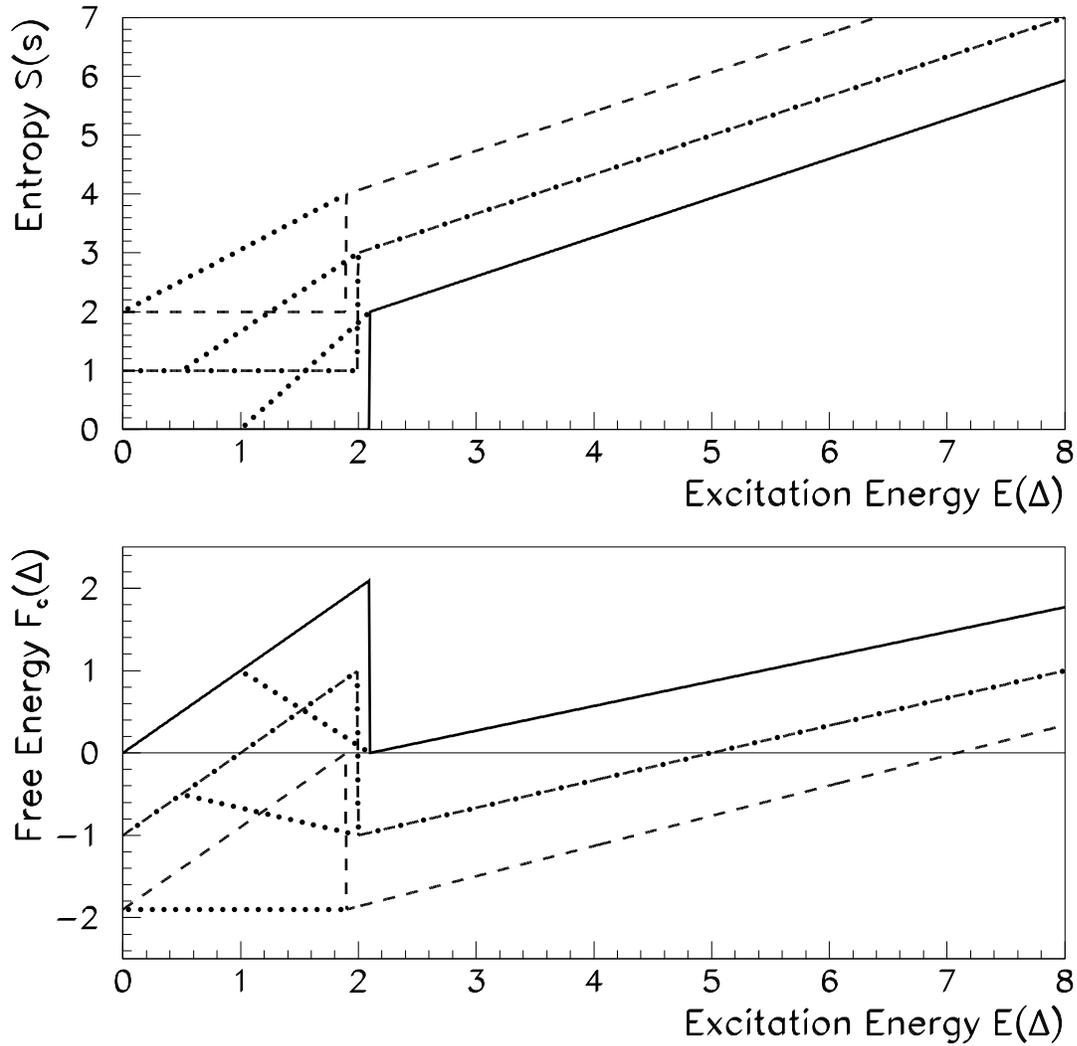}
\caption{Schematic representation of the entropy $S$ in units of the single
particle entropy $s$ (top panel) for even-even (solid line) odd-mass 
(dash-dotted line) and odd-odd (dashed line) nuclei. For the purpose of the 
figure, the steps in entropy are drawn slightly staggered in energy. Lower 
panel: linearized Helmholtz free energy $F_c$ at the critical temperature $T_c$
of even-even, odd-mass and odd-odd nuclei. All energies are measured in units 
of the pairing gap parameter $\Delta$. The dotted lines indicate the situation 
if additional levels are included below the steps in entropy.}
\label{fig:fig2}
\end{figure}

\begin{figure}
\includegraphics[totalheight=20cm,angle=0,bb=0 20 350 730]{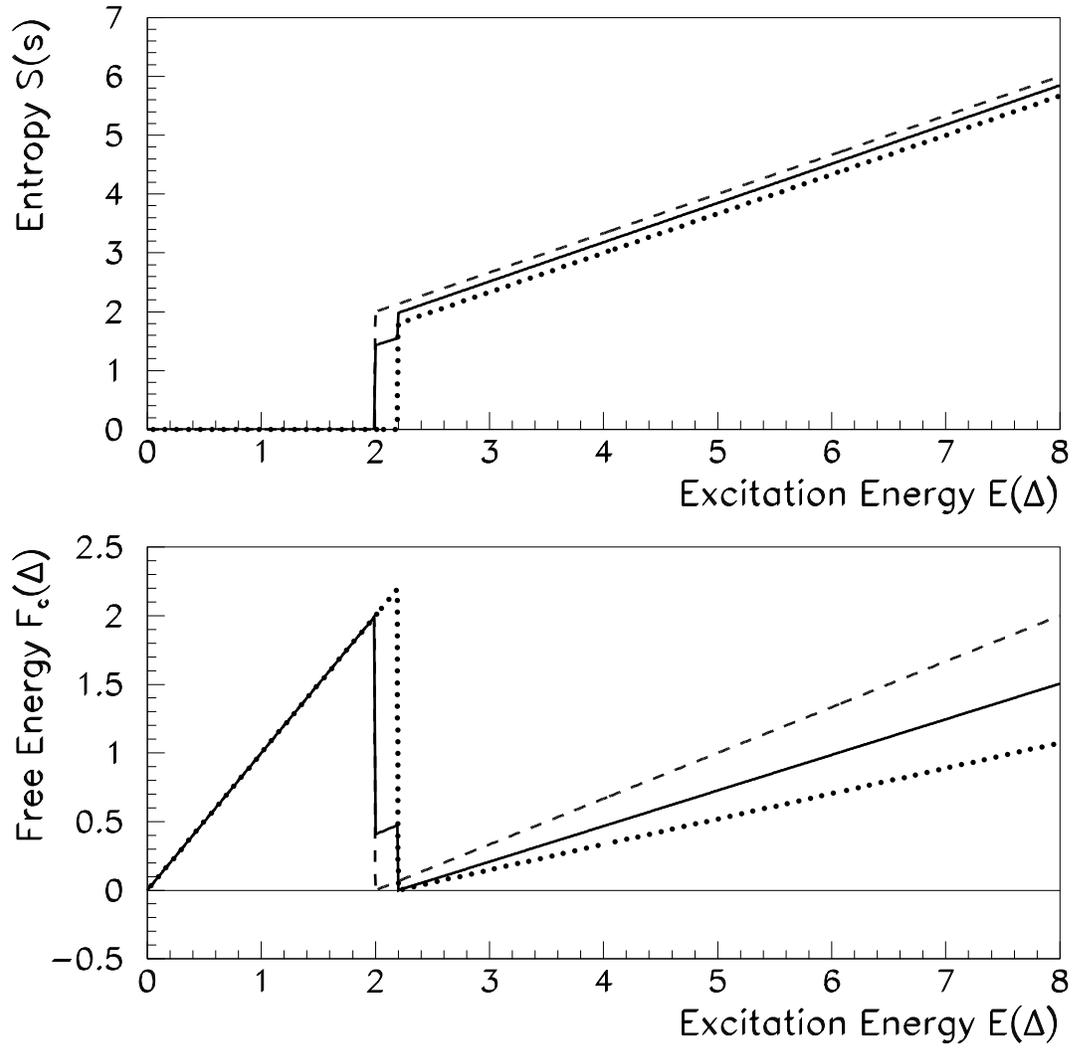}
\caption{Same as previous figure in the case of an even-even nucleus but for 
unequal proton and neutron fluids. Curves are given for the neutron fluid alone
(dashed lines with pairing gap parameter $\Delta$ and single particle entropy 
$s$), the proton fluid alone (dotted lines with $1.1\Delta$ and $0.9s$) and the
composite system (solid lines).}
\label{fig:fig3}
\end{figure}

\begin{figure}
\includegraphics[totalheight=20cm,angle=0,bb=0 20 350 730]{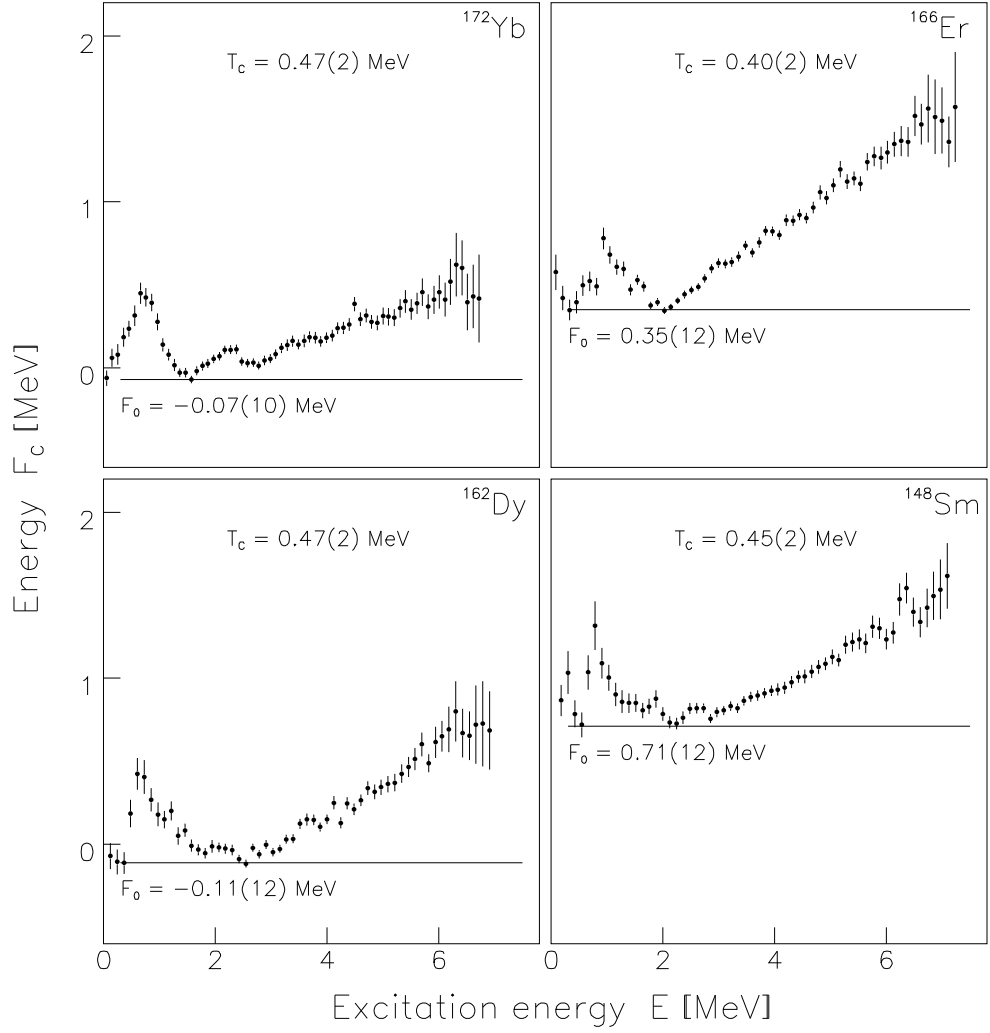}
\caption{Linearized Helmholtz free energy at the critical temperature $T_c$. 
The constant level $F_0$ connecting the two minima is indicated by lines.}
\label{fig:fig4}
\end{figure}

In order to experimentally investigate the behavior for even isotopes, 
linearized free energies $F_c$ for certain temperatures $T_c$ are displayed in 
Fig.\ \ref{fig:fig4}. The data clearly reveal two minima with 
$F_c(E_1)=F_c(E_2)=F_0$, which is due to the general increase in level density 
around $E\sim 2$~MeV, as schematically shown in Fig.\ \ref{fig:fig2}. For all 
nuclei, we obtain $E_1\sim 0$ and $E_2\sim 2$~MeV which compares well with 
$2\Delta$. We interpret the results of Fig.\ \ref{fig:fig4} as the transition 
due to the breaking of the very first nucleon pairs. The deduced critical 
temperatures are $T_c=0.47$, 0.40, 0.47 and 0.45~MeV for $^{172}$Yb, 
$^{166}$Er, $^{162}$Dy, and $^{148}$Sm, respectively.

Recently \cite{SB01}, another method was introduced to determine the critical 
temperatures in the canonical ensemble. Here, the constant temperature level
density formula for the canonical heat capacity $C_V(T)=(1-T/\tau)^{-2}$ was 
fitted to the data in the temperature region of 0--0.4~MeV corresponding to 
excitation energies between 0--2~MeV\@, and the fitted temperature parameter 
$\tau$ was then identified with the critical temperature $T_c$. Since a 
constant temperature level density formula implies a constant linearized 
Helmholtz free energy $F_c(E)$ (provided $\tau=T_c$), this former method is 
almost equivalent to the present method, i.e., of identifying the temperature 
$T_c$ for which the linearized Helmholtz free energy is in average constant. 
Therefore, it is not surprising that the extracted critical temperatures 
$T_c=0.49$, 0.44, 0.49 and 0.45~MeV for the respective nuclei using the older
method \cite{SB01,MG01,SG02} coincide well with the critical temperatures 
presented in this work. However, while the previous method was based on an 
ad-hoc assumption of the applicability of a constant temperature level density 
formula, the present method has a much firmer theoretical foundation.

The height of the free-energy barrier should show a different dependence on the
system size $L$ according to the order of a possible phase transition 
\cite{LK90,LK91}. The barriers deduced from Fig.\ \ref{fig:fig4} yield 
$\Delta F_c\sim 0.5$--0.6~MeV, values which seem not to have any systematic 
dependence on the mass number $A$ within the experimental uncertainties. Even 
with better data, an unambiguous dependence of the barrier height on the system
size would be unlikely when using $A$ as a measure for the parameter $L$ since 
the relevant system size for the very first breaking of Cooper pairs might be
characterized by only a few valence nucleons. Another complicating interference
is that other properties of the nuclear system which might influence the onset
of pair breaking also change with mass number, e.g., deformation, pairing gap, 
and locations of single particle levels around the Fermi surface.

In the schematic model of Fig.\ \ref{fig:fig2}, we would expect a free energy 
barrier of $\Delta F_c=2\Delta\sim 2$~MeV at $E=2\Delta\sim 2$~MeV. However, 
the data are more consistent with the dotted lines of Fig.\ \ref{fig:fig2} 
indicating a smoother behavior around the expected steps due to the existence 
of collective excitations like rotation and $\beta$, $\gamma$, and octupole 
vibrations between 1 and 2~MeV for the even-even nuclei, and due to the 
increasing availability of single-particle orbitals for the odd nucleon in the 
case of odd nuclei. Thus, we expect the centroid of the barrier to be shifted 
down in energy with a corresponding proportional reduction of the barrier 
height, and an inspection of Fig.\ \ref{fig:fig4} indeed shows that the free 
energy barrier is 0.5--0.6~MeV at $\sim 0.6$~MeV excitation energy for the
even-even nuclei. A similar analysis of the odd isotopes is difficult to 
accomplish since there seems not to be any common structures. Here, the 
unpaired valence neutron smears out the effects of the depairing process too 
much to be visible in the present data. However, it has been attempted in Ref.\
\cite{MG01} to interpret the structure in the level density of $^{167}$Er 
around 1~MeV in terms of a first order phase transition.

\begin{figure}
\includegraphics[totalheight=20cm,angle=0,bb=0 20 350 730]{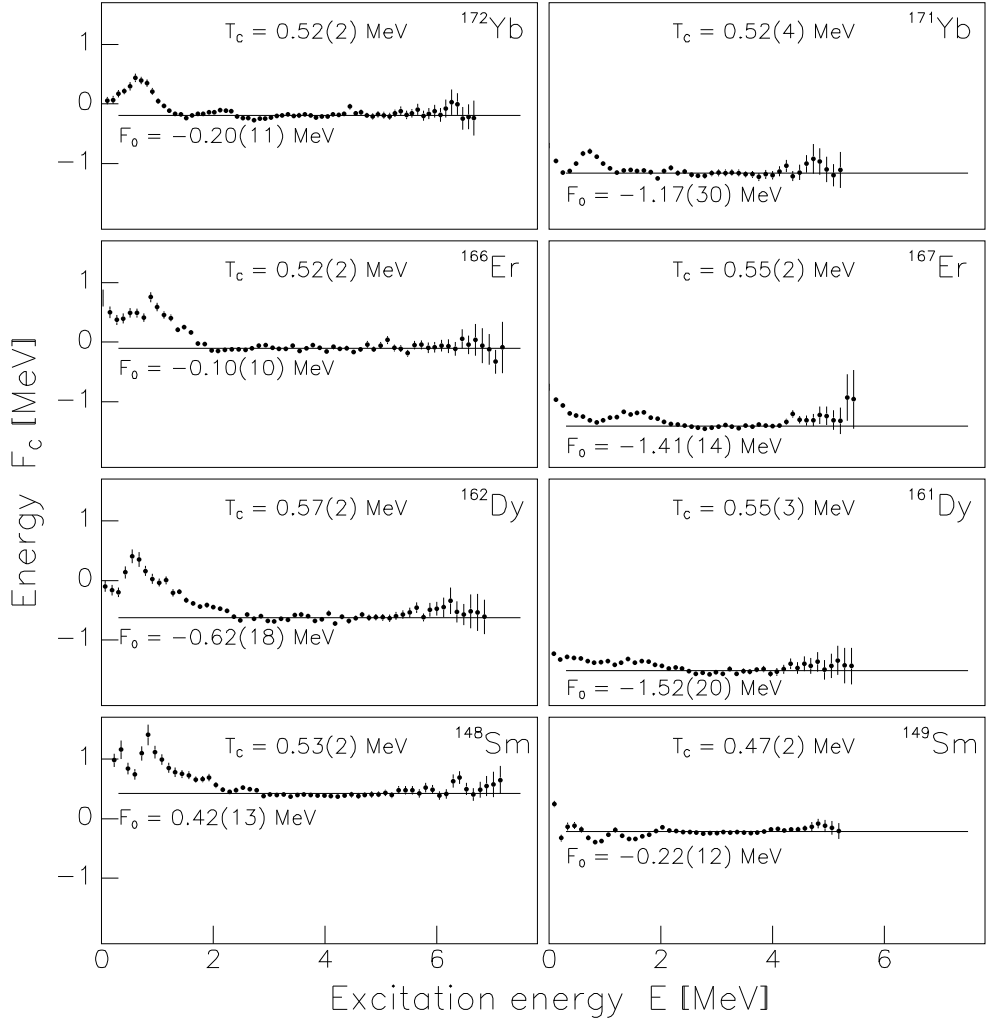}
\caption{Linearized Helmholtz free energy at the critical temperature $T_c$. 
The fitted constant level $F_0$ is indicated by lines.}
\label{fig:fig5}
\end{figure}

The smearing effect is expected to be even more pronounced for the breaking of
additional pairs. Figure \ref{fig:fig5} shows the linearized Helmholtz free 
energy for all eight nuclei investigated, but at slightly higher critical 
temperatures than in Fig.\ \ref{fig:fig4}. The critical temperature $T_c$ is 
found by a least $\chi^2$ fit of a constant value $F_0$ to the experimental 
data. The fit region is from $E=2$~MeV up to 5 and 7~MeV for the odd and even 
isotopes, respectively, giving normalized $\chi^2$ values in the range from 0.5
to 2.5. Here, instead of a double-minimum structure, a continuous "minimum" of 
$F_c$ is displayed for several MeV\@. This observation allows us to conclude 
that the further depairing process cannot under any circumstances be 
interpreted as an abrupt structural change in the nucleus typical for a first 
order phase transition. The constant lines of Fig.\ \ref{fig:fig5} visualize 
how surprisingly well $F_0$ fits the data:\ the deviations are typically less 
than 100~keV\@.\footnote{This fact might also settle the discussion in Ref.\ 
\cite{MG01} and discard the possible interpretation of the many negative 
branches of the microcanonical heat capacity observed in Fig.\ 8 of this 
reference as indicators of separate first-order phase transitions.} The ongoing
breaking of further Cooper pairs overlapping in excitation energies above 2~MeV
is therefore contrary to what is found in the schematic model of Ref.\ 
\cite{BD01}. This is probably due to strong residual interactions in real 
nuclei, like the quadrupole-quadrupole interaction, which were not taken into 
account in the model calculation. Thus, (nearly) all excitation energies above 
2~MeV will energetically match with the costs of breaking nucleon pairs. Here, 
all excess energy goes to the process of breaking pairs. Since the gain in 
entropy $dS$ is proportional to $dE$, the microcanonical temperature, 
$T(E)=(dS/dE)^{-1}$, remains constant as function of excitation energy, and the
level density displays a straight section in the log plot.

At higher excitation energies than measured here, the pairing correlations 
vanish and the system behaves more like a Fermi gas. Here, the free energy will
indicate the closing stage of the depairing process by increasing $F_c$, with 
$F_c>F_0$. However, in this regime also shape transitions and fluctuations as 
well as the melting of the shell structure may play a role and give deviations 
from a simple Fermi gas model with $\rho\propto\exp(2\sqrt{aE})$, $a$ being the
level density parameter. Unfortunately, these very interesting phenomena cannot
be investigated with the present experimental data.

\begin{figure}
\includegraphics[totalheight=20cm,angle=0,bb=0 20 350 730]{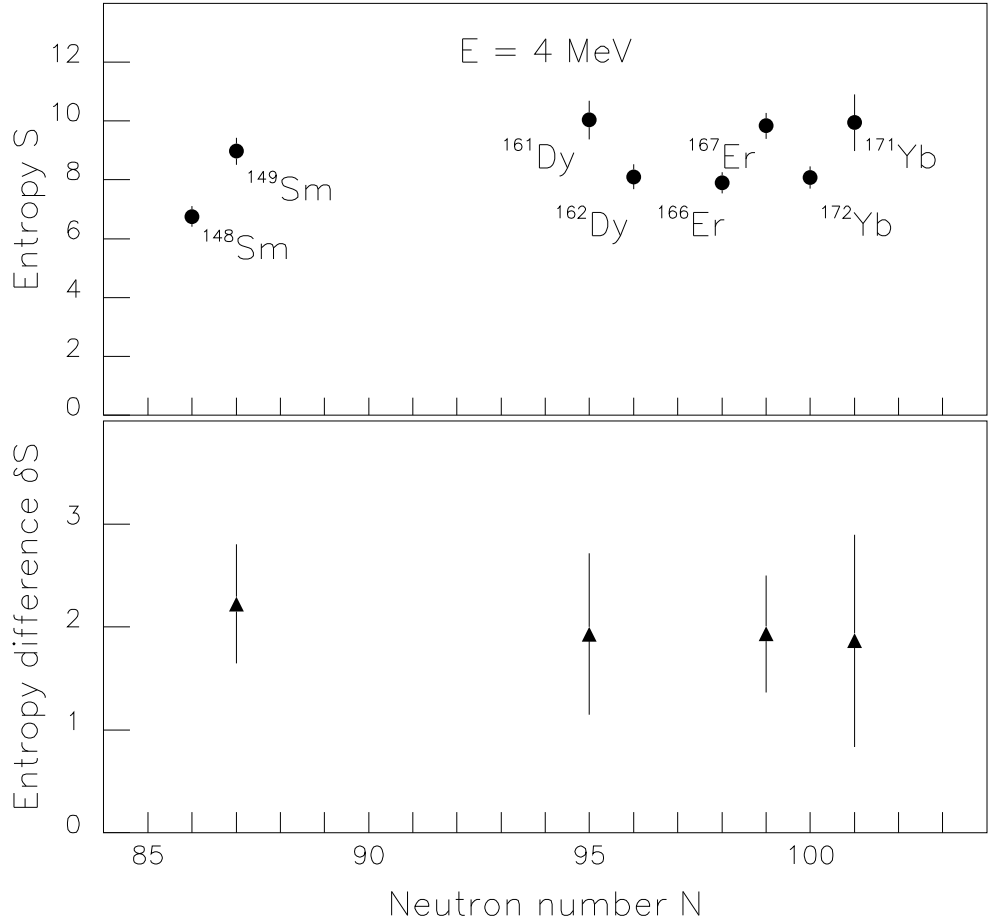}
\caption{Experimental entropy evaluated in the microcanonical ensemble at 
excitation energy $E=4$~MeV and temperature $T_c$. In the lower panel the 
odd-even difference $\delta S=S_{\mathrm{odd}}-S_{\mathrm{even}}$ is displayed 
for the four isotopes.}
\label{fig:fig6}
\end{figure}

The fitted value $F_0$ contains information on the entropy of the system at 
$T_c$ through $S=(E-F_0)/T_c$. In Fig.\ \ref{fig:fig6}, we have compared the 
entropy for the various nuclei at an excitation energy $E=4$~MeV, an energy 
where all nuclei seem to "behave" equally well (see Fig.\ \ref{fig:fig5}). 
Figure \ref{fig:fig6} also shows that the odd mass nuclei display generally 
higher entropy regardless of the mass number $A$ being one higher or lower than
the neighboring even isotope. We also observe that since the $^{148,149}$Sm 
nuclei are not mid-shell nuclei, they show less entropy, reflecting the lower 
single particle level density when approaching the $N=82$ shell gap. By 
evaluating the odd-even $\delta S=S_{\mathrm{odd}}-S_{\mathrm{even}}$, we find 
$\delta S\sim 2$ for all four isotopes, as shown in the lower panel of Fig.\ 
\ref{fig:fig6}. This means that excited holes and particles have the same 
degree of freedom with respect to the even mass nuclei.

\section{Conclusion}

Unique experimental information on level densities for eight rare earth nuclei 
is utilized to extract thermodynamic quantities in the microcanonical ensemble.
The linearized Helmholtz free energy is used to obtain the critical 
temperatures of the depairing process. For a critical temperature just below 
$T_c\sim 0.5$~MeV, we observe a structural transition of even nuclei in the 
$E=0-2$~MeV region due to the breaking of the first nucleon pair. 
Unfortunately, it was not possible to use the development of the barrier height
$\Delta F_c$ with the size of the system $L$ to conclude on the presence of a
thermodynamical phase transition and its order. The critical temperature for 
the melting of other pairs is found at slightly higher temperatures. Here, we 
obtain a surprisingly constant value for the linearized Helmholtz free energy, 
indicating a continuous melting of nucleon Cooper pairs as function of 
excitation energy. The conspicuous absence of a double-minimum structure in 
$F_c$ for this process is at variance with the presence of a first-order phase 
transition in the thermodynamical sense. The entropy difference between odd and
even systems is found to be constant with respect to excitation energy and is 
consistent with the expected values of the single particle entropy in these 
nuclei.

\acknowledgements

Financial support from the Norwegian Research Council (NFR) is gratefully 
acknowledged. Part of this work was performed under the auspices of the U.S. 
Department of Energy by the University of California, Lawrence Livermore 
National Laboratory under Contract No.\ W-7405-ENG-48. Research at Oak Ridge 
National Laboratory was sponsored by the Division of Nuclear Physics, U.S. 
Department of Energy under contract DE-AC05-00OR22725 with UT-Battelle, LLC\@.

\end{multicols}

\clearpage

\clearpage

\clearpage

\clearpage

\clearpage

\clearpage

\end{document}